\documentclass[a4paper]{article}

\usepackage{INTERSPEECH2020}
\usepackage{algorithm}
\usepackage{algorithmic}
\usepackage{epstopdf}
\usepackage{multirow}

\title{Domain-Invariant Speaker Vector Projection by Model-Agnostic Meta-Learning}
\name{Jiawen Kang, Ruiqi Liu, Lantian Li, Yunqi Cai, Dong Wang, Thomas Fang Zheng}
\address{
  Center for Speech and Language Technologies, Tsinghua University, China}
\email{wangdong99@mails.tsinghua.edu.cn}

\begin{document}

\maketitle
\begin{abstract}
Domain generalization remains a critical problem for speaker recognition, even with the state-of-the-art architectures based on deep neural nets. For example, a model trained on reading speech may largely fail when applied to scenarios of singing or movie.
In this paper, we propose a domain-invariant projection to improve the generalizability of speaker vectors. This projection is a simple neural net and is trained following the Model-Agnostic Meta-Learning (MAML) principle, for which the objective is to classify speakers in one domain if it had been updated with speech data in another domain. We tested the proposed method on CNCeleb, a new dataset consisting of single-speaker multi-condition (SSMC) data. The results demonstrated that the MAML-based domain-invariant projection can produce more generalizable speaker vectors, and effectively improve the performance in unseen domains.

\end{abstract}

\section{Introduction}

Speaker recognition (SRE) has gained good performance after decades of research~\cite{hansen2015speaker}.
Most modern SRE approaches are based on \emph{speech embedding}, i.e., representing variable-length speech segments
by fixed-length continuous vectors. This embedding is traditionally derived from statistical models, e.g., the i-vector model~\cite{dehak2011front},
and recently mostly via deep neural nets (DNN)~\cite{ehsan14,li2017deep}, e.g., the x-vector model~\cite{snyder2018xvector,okabe2018attentive}.
The deep embedding models have been significantly improved recently, by employing better architectures~\cite{chung2018voxceleb2,Jung2019raw},
pooling approaches~\cite{okabe2018attentive,Cai2018,Xie19a,Chen2019tied},
training objectives~\cite{ding2018mtgan,Wang2019centroid,bai2019partial,Gao2019improving,Zhou2019deep},
and training schemes~\cite{Li2019boundary,Wang2019phonetic,Stafylakis2019}.
As a result, it has achieved the state-of-the-art (SOTA) performance on several benchmark datasets~\cite{Sadjadi2019},
in particular when combined with the PLDA model~\cite{Ioffe06} for scoring.

In spite of the high performance on existing benchmark datasets, a large performance degradation is often observed
when the deep embedding models are deployed to real applications.
For example, in a preliminary study~\cite{fan2019cn},
we found that a SOTA model trained with the large-scale Voxceleb dataset can achieve great performance on the
SITW evaluation set (less than $2\%$ in equal error rate (EER)), however when applied to a more realistic CNCeleb evaluation set,
the performance degrades to $10\%$-$30\%$ in EER, depending on the genre of the test data.
This degradation should be attributed to the severe domain mismatch caused by
the complex acoustic environments and speaking styles in real-life applications.
Unfortunately, this mismatch is not easy to solve by simply collecting more data,
compared to other speech processing tasks such as automatic speech recognition (ASR).
This is because the speaker property is convolved with other factors in the speech signal.
In order to distinguish the speaker property from other factors,
the training data must contain speech from the same speaker but in different acoustic environments and speaking styles,
i.e., single-speaker and multi-condition (SSMC) data.
In contrast, ASR training requires single-word and multi-condition (SWMC) data.
It is obvious that SSMC data is much more difficult to collect than SWMC data.

A large body of research has been conducted to solve the domain mismatch problem.
The most popular approach is domain adaptation, which adapts
the basic model by a small amount of in-domain data.
Since the embedding model is highly complex, the adaptation is more often performed with the PLDA scoring model.
This adaptation could be supervised or unsupervised.
The supervised approach uses class labels in the target domain,
and adapts PLDA following the Bayesian rule in principle~\cite{villalba2012bayesian,garcia2014supervised}.
The unsupervised approach employs various clustering methods to generate pseudo classes,
and then treats these pseudo classes as true speakers to conduct supervised adaptation~\cite{garcia2014unsupervised}.
%For example, Shon et al.~\cite{shon2017autoencoder} proposed to design a mapping function that
%transfers the data from the source domain to the target domain.
Another approach is domain-invariant training.
Compared to the adaptation approach that targets for better performance in a particular domain,
the domain-invariant training targets for learning domain-insensitive speaker vectors,
and is more amiable to real applications where the conditions may vary in time.
For example, Wang et al.~\cite{wang2018unsupervised} proposed an adversarial loss
that prevents the produced speaker vectors from being domain discriminative.

In this paper, we present a domain-invariant training approach based on meta learning.
Meta learning is a high-level learning strategy with the principle of knowledge sharing and
transferring among tasks~\cite{vilalta2002perspective,vanschoren2018meta}.
In the deep learning regime, early meta-learning approaches learn a training scheme~\cite{bengio1992optimization,andrychowicz2016learning}.
Recently, Finn et al. presented a new Model-Agnostic Meta-Learning (MAML) algorithm~\cite{finn2017model},
which employs data of multiple tasks to learn a model that can be easily adapted to a new task. Borrowing this idea,
we propose a \textbf{robust MAML} algorithm that can learn domain-invariant model directly,
rather than a model that is ready for adaptation as in the standard MAML.
We apply this new algorithm to learn a projection net that
improves the domain invariance of the raw deep speaker vectors (x-vectors in our experiments).
Experimental results on a new SSMC  CNCeleb dataset~\cite{fan2019cn} demonstrated that this approach is promising.

The rest of the paper is organized as follows. Section 2 reviews the basic MAML algorithm, and Section 3 presents the details of
the robust MAML algorithm and the MAML-based project net.
Section 4 presents the experimental result, and Section 5 gives a conclusion of the entire paper.

\section{MAML algorithm}

Meta learning is a long-standing theme in machine learning~\cite{vilalta2002perspective,vanschoren2018meta}.
The central idea of this learning approach is to reuse the knowledge of some \emph{prior} tasks to speed up the learning of a new task.
The knowledge can be either the learned models or the learning strategy (how to learn a task).
For neural models, reusing both types of knowledge is easy.
In the former case, it is known as \emph{multi-task learning} or \emph{transfer learning}~\cite{wang2015transfer},
and in the latter case, a more proper name is \emph{learning to learn}~\cite{bengio1992optimization,andrychowicz2016learning}.

\begin{figure}[htb!]
 \centering
  \includegraphics[width=\linewidth]{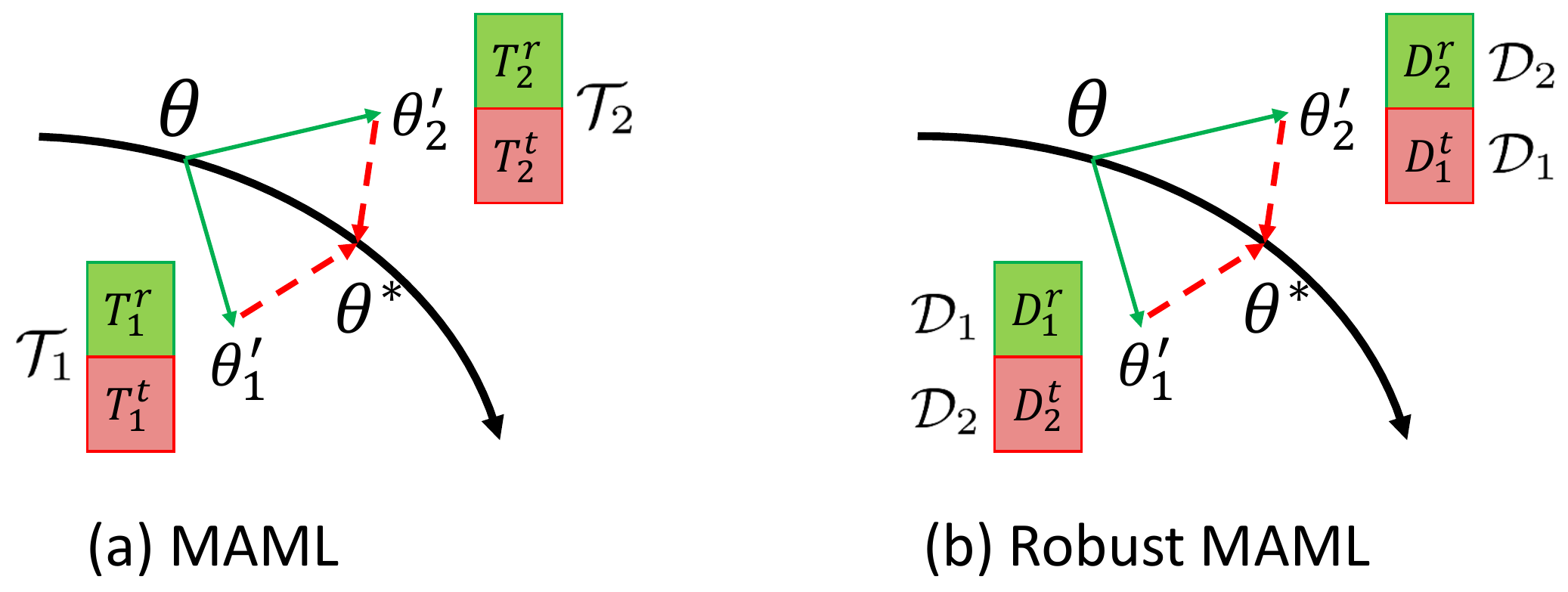}
  \caption{MAML for training a good initial model (a) and robust MAML for training a domain-invariant model (b).
  The green solid line (green block) represents local update, and the red dash line (red block) represents the meta update.}
 \label{fig:bp}
\end{figure}

Model-Agnostic Meta-Learning (MAML)~\cite{finn2017model} is a new meta learning approach.
The concept is shown in Figure~\ref{fig:bp}(a).
In this picture, there are two prior tasks $\mathcal{T}_1$ and $\mathcal{T}_2$,
and the associated training and test data are $(T^r_1,T^t_1)$ and $(T^r_2, T^t_2)$, respectively.
Let $f_{\theta}$ denotes the function of the model parameterized by $\theta$.
For each mini-batch, a small set of training data from $\mathcal{T}_i$, denoted by $m^r_i \sim T^r_i$, is selected.
Using this mini-batch, the gradient $\nabla_{\theta}$ is computed, which is then used to perform a \emph{local update} for the model:

\[
\theta' = \theta - \alpha \nabla_{\theta} \mathcal{L}_{\mathcal{T}_i} (f_\theta; m^r_i),
\]
\noindent where $\mathcal{L}_{\mathcal{T}_i}$ is the loss function of task $\mathcal{T}_i$, and $\alpha$ is the learning rate.
Based on the new parameters $\theta'$, compute the loss on $m^t_i \sim T^t_i$, a mini-batch from the test dataset of the \emph{same} task:

\[
\mathcal{L}_{\mathcal{T}_i}(f_{\theta'}; m^t_i) = \mathcal{L}_{\mathcal{T}_i}(f_{\theta - \alpha \nabla_{\theta} \mathcal{L}_{\mathcal{T}_i} (f_\theta; m^r_i)}; m^t_i).
\]
\noindent This loss is used to compute gradient for model update, which is called the \emph{meta update}.

\[
\theta \gets \theta - \beta \nabla_{\theta} \mathcal{L}_{\mathcal{T}_i}(f_{\theta - \alpha \nabla_{\theta} \mathcal{L}_{\mathcal{T}_i} (f_\theta; m^r_i)}; m^t_i),
\]
\noindent where $\beta$ is the learning rate.
It should be noted that it is the meta update that truly modifies the model parameters.
The local update is just a proxy to compute the gradient for the meta update.

From the training procedure, it can be seen that the goal of MAML is to find an optimal $\theta^*$ at which
\emph{the averaged performance would be best if a simple gradient update had been conducted to adapt the present model to task-specific models}.
In other words, MAML intends to learn a good initial model base on which task-specific models can be
easily obtained by one or a few gradient updates.

\section{MAML-based domain-invariant projection}

\subsection{Robust MAML}

The MAML algorithm discussed in the previous section focuses on a good initial model.
This is valuable for many applications especially those with limited training data.
It can be directly applied to SRE for domain adaptation.
Specifically, we can treat the SRE task on each domain as a particular prior task in the MAML algorithm,
and train an initial model with data from a couple of prior domains.
When deploying to a new domain, a small amount of training data would be sufficient to adapt the initial model to a domain-specific model.
However, this adaptation approach does not meet our original goal of designing a robust model that works well on \emph{any unseen} domain.
We therefore present a robust MAML to deal with the problem.

As shown in Figure~\ref{fig:bp}(b), we have two domains $\mathcal{D}_1$ and $\mathcal{D}_2$, and $\{D_1^r, D_1^t\}$
and $\{D_2^r, D_2^t\}$ are the training and test sets for each of the two domains, respectively.
The MAML training is conducted as usual, but the meta update is based on a mini-batch whose domain is different from the one used for the local update. Put it formally, the local update is conducted by randomly choosing a mini-batch in the $i$-th domain:

\[
  \theta' = \theta - \alpha \nabla_{\theta} \mathcal{L} (f_\theta; m^r_i),
\]
\noindent where $m^r_i \sim D_i^r$. Note that we have omitted the domain dependency in the loss function $\mathcal{L}$ as
all the domains share the same loss function. During the meta update, a mini-batch from the $j$-th domain is selected,
and the loss function is as follows:

\[
  \mathcal{L}(f_{\theta'}; m^t_j) = \mathcal{L}(f_{\theta - \alpha \nabla_{\theta} \mathcal{L} (f_\theta; m^r_i)}; m^t_j),
\]
\noindent where $m^t_j \in D_j^t$, and usually $m^r_i$ and $m^t_i$ are from different domains.
The meta update is then formulated as follows:

\[
\theta \gets \theta - \beta \nabla_{\theta} \mathcal{L}(f_{\theta - \alpha \nabla_{\theta} \mathcal{L} (f_\theta; m^r_i)}; m^t_j).
\]

Choosing different domains for the local update and the meta update is important.
Assume that the model has converged to $\theta^*$,
it is easy to see that updating $\theta^*$ towards \emph{any} domain will result in good performance for \emph{all} domains.
This implies that $\theta^*$ is a stationary point that works well for all the prior domains even without any update.
The idea that training conducted in one domain and evaluated in other domains
aligns to robust optimization~\cite{qian2019robust}. We therefore denote the new algorithm as \textbf{robust MAML}.
Note that a similar idea has been discussed in~\cite{dou2019domain}.

We highlight that with the robust MAML, it is not necessary that the local update and the meta update use the same set of speakers,
although better performance was found if they do. It means that SSMC data is not strictly required for MAML training.
This is a key advantage compared to other domain-robust approaches, for example multi-conditional training.

\begin{figure*}[htb!]
  \centering
  \includegraphics[width=0.90\linewidth]{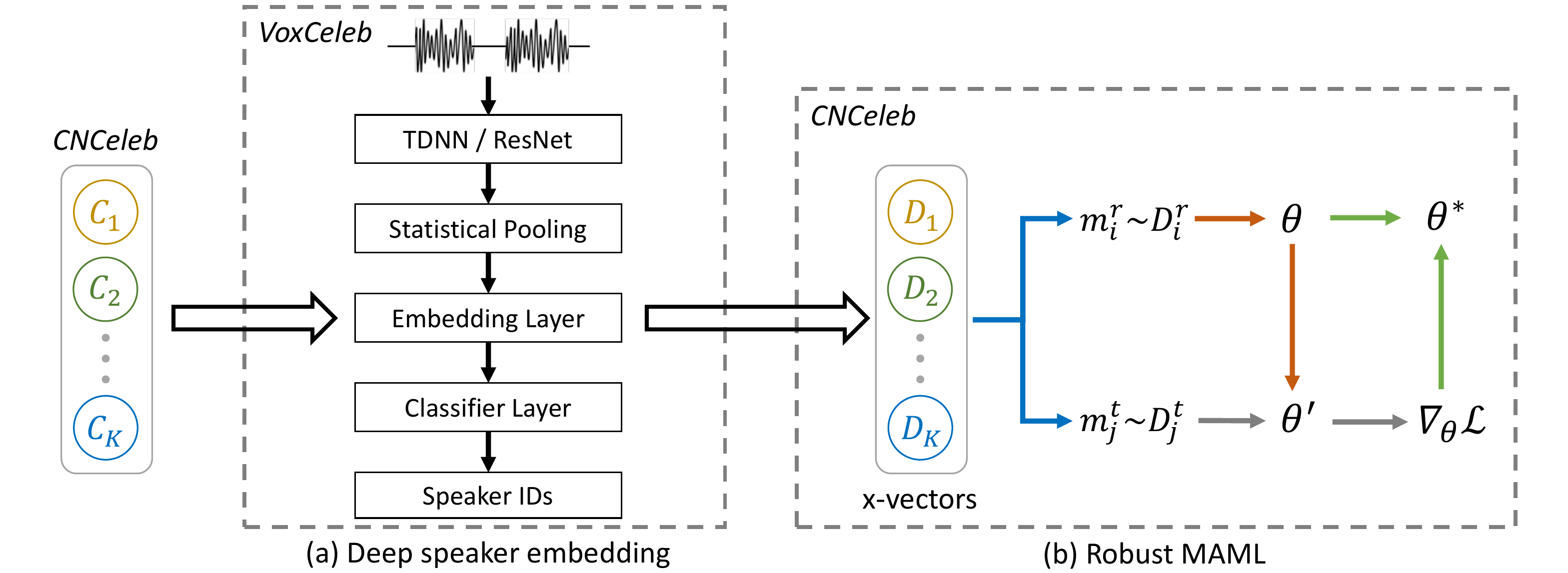}
  \caption{Robust MAML-based domain-invariant projection on deep speaker vectors.}
  \label{fig:structure}
\end{figure*}

\subsection{Domain-invariant projection net}

At the first glance, applying the robust MAML to train domain-invariant speaker embedding models is straightforward.
However, we found it is not applicable in real situations.
A particular problem is domain imbalance: for most of the existing datasets,
a large proportion of the data are clean and reading speech, and only a small amount of data are from other domains.
With this imbalanced data, the minor domains will be overwhelmed by the major domain when conducting the robust MAML training.

To solve this problem, we choose a post-processing scheme.
Firstly, we use a standard large-scale dataset to train the main embedding model, which is the x-vector model in our experiment.
Secondly, we apply the robust MAML algorithm to train an extra projection net that
maps the original x-vectors to a new vector space where domain invariance is improved.
Figure~\ref{fig:structure} illustrates the entire training procedure.

\section{Experiments}

\subsection{Data}

Three datasets were used in our experiments: VoxCeleb~\cite{chung2018voxceleb2,nagrani2017voxceleb}, SITW~\cite{mclaren2016speakers} and
CNCeleb~\cite{fan2019cn}.
More information about these three datasets is presented below.

\noindent \textbf{VoxCeleb}: This is a large-scale audiovisual speaker database collected by the University of Oxford.
The entire database contains $2,000$+ hours of speech signals from $7,000$+ speakers.
It was used to train the x-vector embedding model and the basic LDA/PLDA scoring model.
Data augmentation was applied to improve robustness, with the MUSAN corpus used to generate noisy utterances,
and the room impulse responses (RIRS) corpus used to generate reverberant utterances.

\noindent \textbf{SITW}: This is a standard evaluation dataset excerpted from VoxCeleb1.
In our experiments, the Eval.Core test set, which contains $3,658$ target trials and $718,130$ imposter trials, was used for evaluation.

\noindent \textbf{CNCeleb}: This is a large-scale free speaker recognition dataset collected by Tsinghua University.
It contains more than $130$k utterances from $1,000$ Chinese celebrities.
It covers $11$ diverse domains, and each speaker may have speech samples in multiple domains, therefore is a true SSMC dataset~\cite{fan2019cn}.
The entire dataset was split into two parts: \emph{CNCeleb.Train},
which covers $7$ domains including entertainment, play, vlog, live broadcast, speech, drama and recitation,
and involves $45,370$ utterances from $800$ speakers, was used to train the projection net and the LDA/PLDA scoring model.
\emph{CNCeleb.Eval}, which covers the rest $3$ domains including singing, movie and interview,
and involves $8,729$ utterances from $200$ speakers, was used for evaluation in unseen domains.
%Note that we ignore the domain of advertisement in our experiments due to its small data volume.

\subsection{Embedding models}

We built two x-vector embedding models, one is based on TDNN, and the other is based on ResNet. Both are widely used in SRE research.

\noindent \textbf{TDNN}: This was trained using the Kaldi toolkit~\cite{povey2011kaldi}, following the SITW recipe.
The acoustic features are 40-dimensional FBanks. The main architecture contains three components.
The first component involves $5$ time-delay (TD) layers to learn frame-level speaker features.
The second component computes the mean and standard deviation of the frame-level features.
The third component involves $2$ full-connection (FC) layers and outputs the posterior
probability over the $7,185$ speakers of the VoxCeleb dataset.
Once trained, the $512$-dimensional activations of the penultimate FC layer are read out as the x-vector of the input utterance.

\noindent \textbf{ResNet34}: The ResNet architecture is similar to the TDNN architecture, with two differences:
(1) it uses the ResNet-34 structure to learn frame-level speaker features~\cite{zeinali2019but}; (2) it uses the Additive Angular Marginal
Softmax (AAM-Softmax)~\cite{deng2019arcface} to compute the posterior probabilities over the training speakers.
Again, the $512$-dimensional activations of the penultimate FC layer are read out as the x-vector of the input utterance.

\subsection{Projection networks}

The projection network is designed to transform the raw x-vectors from the embedding model to a new vector space where
domain invariance is improved.

\subsubsection{MAML net}

In our experiments, the projection net involves $3$ FC layers,
and every layer consists of $512$ units.
The loss function is the same as the one used for the embedding model, which is
the standard Softmax for the TDNN model, and the AAM-Softmax for the ResNet34 model.
The projection net can be trained with CNCeleb.Train, by using the robust MAML algorithm.
We denote the projection net trained in this way as a \emph{MAML net}.
Once the training is completed, the domain-invariant x-vectors can be obtained from the penultimate FC layer.

\subsubsection{MCT net}

Using the same architecture and the same loss function as the MAML net, we can train the projection net
using the regular training scheme. Since the training data (CNCeleb.Train) is SSMC,
this is essentially a multi-conditional training (MCT). We call the MCT-trained projection net
as a \emph{MCT net}. Similar to the MAML net, the MCT net can improve the domain invariance of speaker vectors.
However, the MCT net purely relies on the SSMC data, while the MAML net relies both the SSMC data and the robust training scheme.

\subsection{Baseline results}

We firstly test the performance of the baseline systems, i.e., both the embedding model and the LDA/PLDA scoring
model are trained with VoxCeleb, without any additional speaker vector projection.
We test the performance on the SITW Eval.Core test set and also the CNCeleb.Eval test sets,
including three individual domains (singing, movie and interview).
%the entire set and
The results in terms of equal error rate (EER) are reported in Table~\ref{tab:base},
where the results with two scoring methods, cosine scoring and LDA/PLDA scoring, are reported, respectively.
For the LDA/PLDA scoring, the x-vectors are firstly pre-processed by LDA, and then are used to compute scores by PLDA.
The dimensionality of the LDA projection was set to $128$ in our experiments.
%, which delivered the best performance

%\begin{table}[htb!]
%  \caption{Performance (EER\%) of the baseline systems.}
%  \label{tab:base}
%  \centering
%  \scalebox{0.85}{
%  \begin{tabular}{l|cc|cc}
%  \hline
%                                  &\multicolumn{2}{c|}{TDNN} & \multicolumn{2}{|c}{ResNet34} \\
%  \hline
%                                  & Cosine     &  LDA/PLDA     & Cosine & LDA/PLDA \\
%  \hline
%        SITW.Eval.Core            &   5.139        &   2.433           &    3.226    & 1.958\\
%        CNCeleb.Eval              &   18.10        &   12.61          &   17.24     &  12.17\\
%        CNCeleb-Singing    &   29.95        &   26.88           &   28.47     &  27.18\\
%        CNCeleb-Movie      &   26.09        &   20.24           &    25.19    &  21.29\\
%        CNCeleb-Interview  &   19.68        &   15.97           &    19.23    &  15.47\\
%  \hline
%  \end{tabular}}
%\end{table}
%

\begin{table}[htb!]
  \caption{Performance (EER\%) of the baseline systems.}
  \label{tab:base}
  \centering
  \scalebox{0.85}{
  \begin{tabular}{lcccc}
  \hline
    \multirow{3}{*}{Test Set} & \multicolumn{2}{c}{TDNN} & \multicolumn{2}{c}{ResNet34} \\
                                \cmidrule(r){2-3} \cmidrule(r){4-5}
                              &   Cosine   &  LDA/PLDA   &   Cosine   &  LDA/PLDA     \\
                                \cmidrule(r){1-1} \cmidrule(r){2-3} \cmidrule(r){4-5}
        SITW.Eval.Core        &   5.139    &   2.433     &   3.226    & 1.968\\
%       CNCeleb.Eval          &   18.10    &   12.61     &   17.24    &  12.17\\
        CNC.Eval.Singing       &   29.95    &   26.88     &   28.47    &  27.18\\
        CNC.Eval.Movie         &   26.09    &   20.24     &   25.19    &  21.29\\
        CNC.Eval.Interview     &   19.68    &   15.97     &   19.23    &  15.47\\
  \hline
  \end{tabular}}
\end{table}

\subsection{Results with domain-invariant projection}

In this experiment, we test the MAML net and MCT net on the CNCeleb.Eval test sets, where the domains
are never seen in the training data of the embedding models, the LDA/PLDA scoring models and the projection networks.

\subsubsection{Performance with cosine scoring}

First look at the performance with the cosine scoring.
Since the complex back-end scoring models are not used, we can evaluate the true quality of the speaker vectors.
The results on CNCeleb.Eval test sets are presented in Table~\ref{tab:cosine}. For
convenience, the baseline results are presented in the `Base' columns.

It can be observed that in all the test conditions,
both the MCT net and MAML net can substantially improved the system performance,
and the MAML net offers more significant improvement. This demonstrates that the multi-conditional training scheme
with SSMC data is an effective way to improve domain invariance (MCT vs. Ori), and the robust MAML training
scheme can provide additional and substantial contribution (MAML vs. MCT).

\begin{table}[htb!]
  \caption{Performance (EER\%) with cosine scoring.}
  \label{tab:cosine}
  \centering
  \scalebox{0.88}{
  \begin{tabular}{lcccccc}
  \hline
   \textbf{Cosine} & \multicolumn{3}{c}{TDNN} & \multicolumn{3}{c}{ResNet34} \\
   \cmidrule(r){1-1} \cmidrule(r){2-4} \cmidrule(r){5-7}
        Domain          & Base     & MCT     & MAML     & Base     & MCT     & MAML     \\
   \cmidrule(r){1-1} \cmidrule(r){2-4} \cmidrule(r){5-7}
        Singing         & 29.95   & 30.85   & 29.86    & 28.47   & 28.40   & \textbf{27.08}    \\
        Movie           & 26.09   & 25.46   & 24.27    & 25.19   & 24.92   & \textbf{24.21}    \\
        Interview       & 19.68   & 17.51  & \textbf{16.82}    & 19.23   & 16.92   & 16.87    \\ \hline
    \end{tabular}}
\end{table}

\begin{figure}[htb!]
 \centering
  \includegraphics[width=\linewidth]{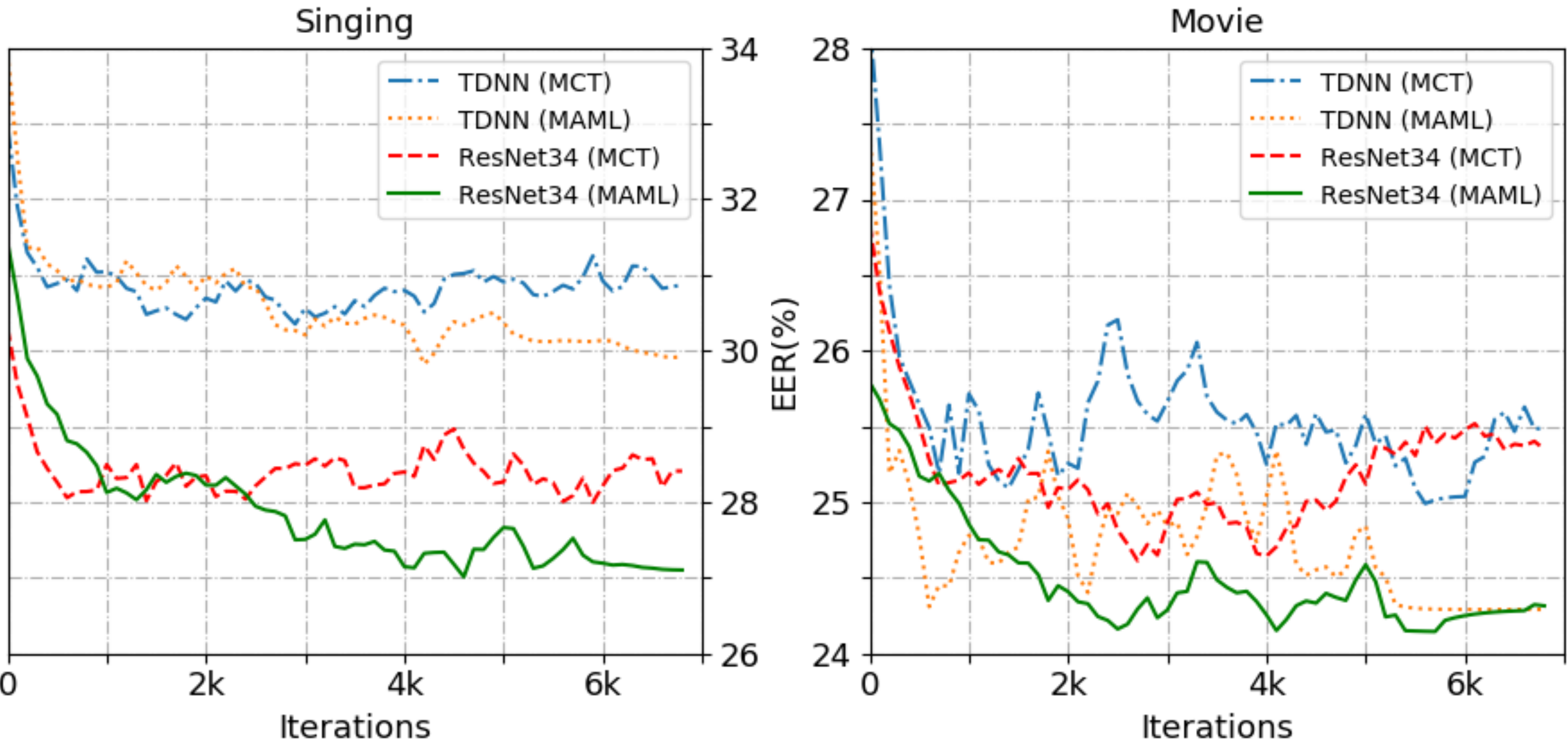}
  \caption{Performance (EER\%) on two unseen domains during the training process of MCT net and MAML net.}
 \label{fig:tp}
\end{figure}

To give a better comparison between the MCT scheme and the MAML scheme, Figure~\ref{fig:tp} shows their performance
along with the training process.
It can be seen that for both the MCT net and the MAML net, the EER continuously reduces.
Moreover, the MAML net delivers better performance than the MCT net.
Compared the two ResNet34 curves in both pictures, we observe that the MCT net seems overfitting after $4$k iterations,
while the MAML net is generally healthy.

\subsubsection{Performance with LDA/PLDA scoring}

We finally test the performance with LDA/PLDA scoring.
In this experiments, all the LDA/PLDA scoring models were trained with CNCeleb.Train.
For the baseline system, we retrained the LDA/PLDA models with the raw x-vectors from the embedding model.
For the MCT and MAML systems, the LDA/PLDA models were trained with the x-vectors produced by the MCT net and MAML net.
Again, the dimensionality of the LDA projection was set to $128$.
Since the training data (CNCeleb.Train) is SSMC, this training scheme of LDA/PLDA is essentially a multi-conditional training (MCT),
and therefore should be robust against domain variance in a way.
%We call these LDA/PLDA models as \emph{MCT-LDA/PLDA}.

The results are shown in Table~\ref{tab:plda}. Firstly, compared to the results in Table~\ref{tab:base},
it can be observed that the MCT-based LDA/PLDA models offer dramatic performance improvement on the test data.
Secondly, the contributions of the MCT net and the MAML net are both marginal. This is not surprising,
as the x-vectors for LDA/PLDA training and MCT/MAML nets training are duplicated.
From the perspective of practitioners, this is a good thing as it implies that domain invariance could be largely
attained by retraining the LDA/PLDA scoring model with SSMC data.
On the other hand, it suggests that the MAML training scheme should be extended to the scoring model,
otherwise its contribution on the speaker vectors will be diminished.

\begin{table}[htb!]
  \caption{Performance (EER\%) with LDA/PLDA scoring.}
  \label{tab:plda}
  \centering
  \scalebox{0.86}{
  \begin{tabular}{lcccccc}
  \hline
   \textbf{LDA/PLDA} & \multicolumn{3}{c}{TDNN} & \multicolumn{3}{c}{ResNet34} \\
   \cmidrule(r){1-1} \cmidrule(r){2-4} \cmidrule(r){5-7}
        Domain          & Base    & MCT     & MAML     & Base    & MCT     & MAML     \\
   \cmidrule(r){1-1} \cmidrule(r){2-4} \cmidrule(r){5-7}
        Singing         & 25.67   & 25.50   & 25.35    & 23.83   & 23.66   & \textbf{23.53}    \\
        Movie           & 19.63   & 18.74   & 18.85    & 18.19   & \textbf{17.75}   & \textbf{17.75}    \\
        Interview       & 13.63   & 13.47   & 13.58    & 12.05   & \textbf{11.85}   & \textbf{11.85}    \\
   \hline
  \end{tabular}}
\end{table}

%\subsection{Visualization}

%T-SNE~\cite{maaten2008visualizing} is used to visualize the speaker vectors in the $2$-dimensional space.
%In Figure~xxx, we choose speaker vectors from $6$ speakers and each speaker covers $2$-$3$ domains,
%and then draw the speaker vectors produced with and without DIP net.
%It can be observed that xxx

\section{Conclusions}

This paper proposed a domain-invariant projection to improve the generalizability of speaker vectors.
We presented a robust MAML algorithm to train the projection net, which promotes domain invariance not only
by the SSMC data, but also by the robust training scheme.
Experimental results on the CNCeleb dataset demonstrated that the speaker vectors produced by MAML-based projection
are more domain-invariant compared to the raw x-vectors and the speaker vectors produced by a multi-conditional trained projection.
This leads to significant performance improvement with cosine scoring.
However, when the scoring model is an LDA/PLDA that was trained with SSMC data, the contribution of the projection net seems marginal.
Future work will investigate the MAML-based training for the LDA/PLDA scoring model, and investigate the light-weighted MAML-based adaptation.

%\section{Acknowledgements}
%This work was supported by the National Natural Science Foundation of China No. 61633013,
%and the Postdoctoral Science Foundation of China No. 2018M640133.
%Jiawen Kang and Ruiqi Liu are joint first authors.

\newpage

\bibliographystyle{IEEEtran}
\bibliography{ref}

\end{document}